\documentclass[showpacs,showkeys,superscriptaddress,twocolumn,floatfix,nofootinbib,secnumarabic]{revtex4-1}
\pdfoutput=1
\usepackage[pdftex]{graphicx}
\usepackage[utf8]{inputenc}
\usepackage{amssymb,amsmath,bm,mathrsfs}
\usepackage{color}
\usepackage{substr}
\definecolor{vdarkblue}{rgb}{0.9,0,0.5}
\usepackage[pdfpagelabels,hyperfootnotes=true,colorlinks,citecolor=blue,urlcolor=blue]{hyperref}
\let\footN\footnote 
\def\footnote#1{\footN{{\color{darkgreen} #1}}}
\let\footTEXT\footnotetext
\def\footnotetext#1{\footTEXT{{\color{darkgreen} #1}}}

\def\RR{\mathbb R}

\def\ie{\mbox{\em i.e.}{} }\def\cf{\mbox{\em cf.}{} }
\def\ff{\mbox{\em ff}{} }
\def\eg{\mbox{\em e.g.}{} }
\def\etc.{\mbox{\em etc.}{} }\def\etal.{\mbox{\em et al.}{} }
\def\Eq.{\mbox{Eq.}{} }\def\DEq.{\mbox{DEq.}{} }
\def\Eqs.{\mbox{Eqs.}{} }\def\DEqs.{\mbox{DEqs.}{} }

\def\cal{\mathcal}\def\e{{\rm e}}

\def\d{{\rm d}}\def\k{\vec{k}\,}
\def\x{\vec{x}\,}
\def\Ft{F}
\def\ft{f}
\def\s{{\sigma}}
\def\la_#1{\lambda_{\rm #1}}\def\F_#1{\Ft_{\rm #1}}\def\f_#1{\ft_{\rm #1}}
\def\n_#1{n_{\rm #1}}\def\N_#1{N_{\rm #1}}

\def\msbx#1{\scalebox{0.9}{\raisebox{0.06mm}{$\,\displaystyle #1$}}}
\def\pbox#1{\begin{picture}(0,0)\put(0,0){{#1}}\end{picture}\hphantom{{#1}}}
\def\sfrac#1#2{{\msbx{\frac{\raisebox{-0.5mm}{$\displaystyle #1$}}{#2}}}}

\definecolor{lightgreen}{rgb}{0.45,0.75,0.15}
\definecolor{darkgreen}{rgb}{0,0.5,0.4}
\definecolor{brown}{rgb}{1.,0.65,0}
\definecolor{lightblue}{rgb}{0.4,0.4,1}

\begin{document}
\title{The  Fate of  Weakly  Bound Light  Nuclei  in Central  Collider
  Experiments:\\ a Challenge in Favor  of a Late Continuous Decoupling
  Mechanism}

\author{J\"o{}rn
  Knoll}\thanks{e-mail:j.knoll@gsi.de} \affiliation{GSI
  Helmholtzzentrum f\"ur Schwerionenforschung GmbH, Planckstr. 1, 64291
  Darmstadt, Germany}

\begin{abstract}
  Arguments are presented  that the reaction products  of central high
  energy nuclear collisions up to  collider energies can rigorously be
  interpreted in terms  of a continuous decoupling  mechanism based on
  continuous  equations  of  motion.    The  various  aspects  of  the
  collision  dynamics  are  investigate  in terms  of  the  individual
  decoupling  processes.   Thereby  each observed  particle  decouples
  during its own  temporal decoupling window.  This  includes a ``very
  late''  decoupling of  the  faintly bound  Hypertritons observed  in
  recent ALICE experiments.  The success of the strategy is based upon
  200 years  old wisdom and leads  to a revised interpretation  of the
  entire decoupling process.
\end{abstract}
\date{\today}%
\maketitle%

\section{General Remarks}
\label{GeneralRemarks}%

Since  the early  days of  high energy  nuclear collisions,  \ie since
about  4 decades,  the abundances  of produced  hadrons and  composite
nuclei were experimentally recorded over a wide range of beam energies
up  to collider  energies at  RHIC  and LHC,  the latter  in order  to
explore    the     QCD    phase-transition    dynamics,     \cf    \eg
Refs.\,\cite{Barz:1988md,Barz:1990qy}.
The recently observed production rates are in part covering up to nine
orders of magnitude for the  various reaction products, \cf the recent
review  \cite{arXiv:1710.09425} compiled  and  published  by my  local
colleagues.   However,  much  to  the  surprise  of  many  colleagues,
including me: since the late 80$^{th}$ the ``statistical hadronization
models'' showed  an amazing stamina.   They provide excellent  fits to
this  wealth of  data, \cf~Ref.\,\cite[Fig.\,2]{arXiv:1710.09425}  and
further   references  therein,   with  solely   two  parameters:   the
temperature  and the  baryon-chemical  potential $(T,\mu_{\rm  B})_0$,
thereby assuming instantaneous global equilibration.
Particularly  intricate and  so  far unsettled  seemed the  production
mechanism of very loosely bound nuclei,  such as deuterons or the even
by   far   more    faintly   bound   Hypertritons\footnote{Hypertriton
$[_{\,3}^\Lambda{\rm H}]$ is a  deuteron with a $\Lambda$-Hyperon halo
with  solely  130\,keV  binding  energy and  an  rms-radius  of  about
10\,fm\,\cite{Juric:1973zq}.},  the latter  with halo  tails of  their
wave-functions  extending further  than  the expected  system size  at
\mbox{$T_0=160$\,MeV},    \cf    the    question   raised    in    the
\mbox{``Outlook'' section}  of Ref.~\cite{arXiv:1710.09425}.  Clearly,
such loosely  bound nuclei are not  expected to decouple at  such high
densities and temperatures\,\mbox{\cite{Sinyukov:2002if, Knoll:2008sc,
    Friman:2011zz:III.Sec.5.8}}.\vspace*{3mm}

\newcounter{freeze}%
Let me start with some principle clarifications:
\begin{list}{\bf\arabic{freeze}:}
  {\usecounter{freeze}\labelsep0.8mm\leftmargin3.8mm\itemsep-0.5mm}%
  \item\label{Nota} The laws  of physics are ruled  by {\em continuous
    equations  of  motion}\,\footnote{This  classifies and  thus  bans
  discontinuous methods  \`a la Cooper-Frye, or  \mbox{coalescence} as
  {\em  inappropriate  theoretical  tools}\,  \cf\!\footnotemark[3].},
    irrespective  in   which  kind  of  approxi\-mation   scheme  ever
    calculated, with  corresponding stochastic interpretations  in the
    quantum case.   For the following  it is furthermore  important to
    clarify  the difference  between  the here  used  notions of  {\em
      FREEZING-IN} and \mbox{\em DECOUPLING}\,.
\item\label{Freeze-in}  {\bf Freezing-in}  defines the  possibly early
  situations, from where on  certain {\em in-medium properties} become
  about   stationary  and   finally   approximately   equate  to   the
  experimentally  observed values.   Such  conclusions  can solely  be
  stated on the  basis of model considerations.
\item
  {\bf Decoupling} is a process which occurs, when matter is subjected
  to any type of  structural changes.  This concerns phase-transitions
  as  well  as  the  here  addressed  release  of  particles  from  an
  interacting  medium.   It  thus   defines  the  continuous  physical
  process,  by   which  the   particles  finally  decouple   from  the
  corresponding   previous   phase.     These   processes   are   {\em
    individual}\,, since  they rely \eg on  interaction cross sections
  \etc.   \\  \hspace*{3mm}The  here addressed  process  concerns  the
  decoupling  from  the  interaction  zone, such  that  the  decoupled
  particles from  then on can  leave unperturbed as free  particles ad
  infinitum!
\item\label{freeze-out}  The  ubiquitously   used  {\bf  instantaneous
  Freeze-out concepts}\,\footnote{Discontinuous  freeze-out  schemes
    are  implicitly used,  if  calculated in-medium  spectra are  {\em
      equated}  with the  asymptotically measured  spectra (they  thus
    bypass the  decoupling process).} tentatively mingle the two above
  aspects  and  may  seriously  misguide conclusions  about  the  here
  addressed decoupling mechanism.
\item\label{Item-Adiabaticity}   These    colliding   systems   evolve
  completely adiabatically without external  influence. This allows to
  describe their evolutions in terms of {hydrodynamics} assuming local
  thermo-chemical equilibrium along every flowline with an appropriate
  coarse   graining  procedure\footnote{Spatially   expanding  systems
  require a volume growth adapted coarse graining procedure, such that
  \eg  the mean  number of  baryons  per fluid  cell is  approximately
  conserved.}.  Collecting the properties of  all fluid cells with the
  same local hydrodynamic properties  then defines the Grand Canonical
  equilibrium    ensemble     of    the    measured     spectra    and
  abundances.
\end{list}

\section{The continuous Decoupling picture}
\label{DecPicture}%

The aforementioned  deficiency \ref{freeze-out} can be  avoided by the
here proposed concept of continuous  decoupling. It is common textbook
wisdom \eg in {cosmo\-logy}\,\cite{Mukhanov:2005sc}, here presented in
the Boltzmann Equation (BE)  approach\footnote{Generalizations of the
decoupling  scheme  to  include  non-local  and  quantum  effects  are
possible  \cf  \eg  my   2008  paper  \cite{Knoll:2008sc}.}   assuming
perturbative decoupling.   Furthermore a  simplified form is  used for
the  here addressed  bulk decoupling  assuming spatial  homogeneity in
local proper time $\tau$.

Here we review  the {\em individual properties} of  each particle type
with mass  $m$.  The individual  detector yields valid for  bosons and
fermions       then      result       to\,\mbox{\cite{Sinyukov:2002if,
    Friman:2011zz:III.Sec.5.8,     Knoll:2008sc,     Mukhanov:2005sc}}
\unitlength1mm

\begin{figure}[b]
  \caption{}  
  \label{fig-Pdec}%
  \begin{picture}(80,23)
    \put(-3.2,-1){\includegraphics[width=0.485\textwidth]{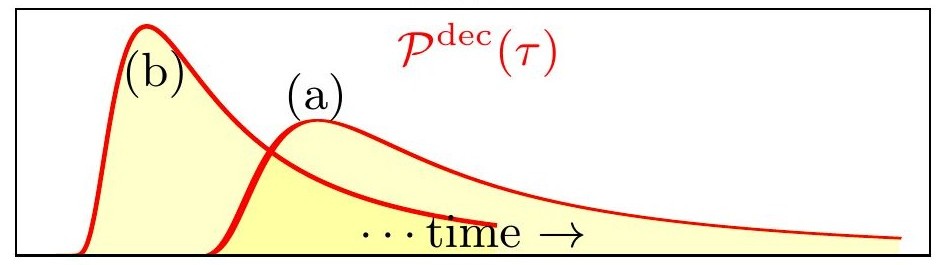}}
  \end{picture}
  \begin{trivlist}\item
    {\bf Schematic plot of the  decoupling probability} defined in \cf
    Eq.\,(\ref{P^dec}) for  two different particles with  particle (a)
    coupling stronger  to the medium  than particle (b),  \cf estimate
    (\ref{t_max}) below.
  \end{trivlist}
\end{figure}%
\begin{subequations}
  \label{dec-rate}
  \begin{align}
    \hspace*{-7mm}
     N^{\rm dec}(\tau)
    &=\int\!\!\!
    \sfrac {\d^3x\d^3k}{(2\pi\hbar)^3}\!\int_{\tau}^\infty\!\!\!\d\tau'
    F(\x,\k,\tau'){\cal  P}^{\rm  dec}(\tau'),
    \hspace*{-3mm}
    \label{dec-rate-a}
    \\
    \label{P^dec}
    \hspace*{-7.6mm} {\cal  P}^{\rm  dec}(\tau)
    &=\Gamma(\tau)
    \exp\Big\{\textstyle \!\!-\!\int\limits_\tau^\infty\!
        \Gamma(\tau')\d\tau'\Big\},
        \hspace*{-3mm}
        \\
        \label{Gamma(tau)}
    \Gamma(\tau)&=\big<\sigma_{\rm tot}|v_{\rm rel}|n(\tau)\big>_{\rm BE}
    \\
    \label{P^dec-unity}
    \mbox{with}~~
    &~~~~\int_{-\infty}^\infty\d\tau\, {\cal  P}^{\rm  dec}(\tau)\equiv 1.
  \end{align}
\end{subequations}

Thereby \pbox{$F(\x,\k,\tau)$}  and ${\cal P}_{\rm  dec}(\tau)$ denote
each particle's local phase-space occupation and individual decoupling
probability\,\cite{Knoll:2008sc},  \cf   Fig.\,\ref{fig-Pdec},  retro
respectively determined by a complete ``ad infinitum'' solution of the
BE analyzed in  the adiabatic coarse graining  context of introductory
Note\;\ref{Item-Adiabaticity}. The local  damping rates $\Gamma(\tau)$
depend on  total cross-sections  of the  concerned particles  with the
surrounding  medium  with  {density} $n(\tau)$  at  relative  velocity
$v_{\rm  rel}$, \cf  (\ref{Gamma(tau)}).  Furthermore each  decoupling
probability  (\ref{P^dec})  integrates  to  unity  (\ref{P^dec-unity})
along  each   fluid-cell's  future   path  assuring   particle  number
conservation.

The  individual BE  rates \mbox{$F(\x,\k,\tau)\times\Gamma(\tau)$}  in
Eq.\;(\ref{dec-rate-a}) describe  the local  creation of  the observed
particles at phase-space point  \pbox{$(\x,\k)$} and local time $\tau$
due to transport processes with  the surrounding medium.  The straight
escape paths  to {\em  asymptotia}\, are assumed  to proceed  in close
vicinity  of  the corresponding  local  fluid  cells. Along  them  the
exponential  factor  (\ref{P^dec})   then  determines  the  particles'
individual \mbox{\em  survival} {\em probabilities}, in  future not to
be  ``kicked''  off  this  mode   along  their  escape  paths  through
interactions with the \mbox{surrounding} medium.
\unitlength0.1\textwidth%
\begin{figure}[b]
  \caption{}
  \label{fig-kappa-adiabat}%
  \begin{picture}(4.6,1.45)
    \put(-0.1,-0.11){\includegraphics[width=0.467\textwidth]{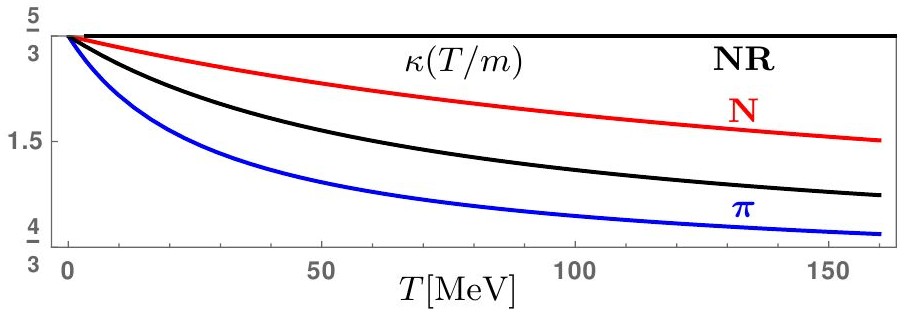}}
  \end{picture}
  \begin{trivlist}\item
    \mbox{\bf            Local           adiabatic            exponent
      $\boldsymbol\kappa$($\boldsymbol T\boldsymbol  /\boldsymbol m$)}
    for       adiabatic       volume       evolution,       \mbox{$T\,
      V(T)^{\kappa(T)-1}=const$},   \cf    Eq.~(\ref{kappa(T)}),   for
    monomer  gases  in  Boltzmann-statistic  conserving  corresponding
    particle         numbers         for:         NR         particles
    \mbox{($m\!\rightarrow\!\infty$)},   relativistic  \mbox{nucleons}
    (N), particles  with intermediate  mass of 350\,MeV  (middle black
    line)    and    pions    ($\pi$).     For    massless    particles
    \mbox{$\kappa=4/3$}.
  \end{trivlist}%
\end{figure}%

Omitting the $\tau$-dependencies  the adiabatically\;evolved densities
$n(T,\mu)$ of each relativistic \mbox{particle} become
\begin{subequations}%
  \label{EQ_properties}%
  \begin{align}
    \label{N(tau)}%
    \hspace*{-5mm}
    n(T,\mu)&=\!\int\!\!\!\sfrac{\d^3k}{(2\pi\hbar)^3}\, F_{\rm eq}
    =\,\underbrace{\!
      \Big(\!\!\sfrac{mT}{2\pi\hbar^2}\!\Big)^{\!3/2} f_{\rm eq}}_{\rm NR}
    \,\lambda_{\rm rel}(T/m)
    \hspace*{-0.7mm}
    \\
    \hspace*{-2.5mm}
    \label{lambda}
    \mbox{with}\hspace*{3mm}
    f_{\rm eq}&=\e^{\,\mu_{\rm NR}^{~}/T}\!,~
    \lambda_{\rm rel}(T/m)\approx 1 +\!\sfrac{2T}{m}+\!\sfrac{T^2}{2m^2},\!
    \\
    \hspace*{-3.5mm}
    F^{\,\rm  eq}&=\exp[\{\mu-(m^2+\k^2)^{1/2}\}/T]
    \label{F^eq}
    \\
    p(T,\mu)&=n(T,\mu)\,T~~~\mbox{(even relativistic)},
    \label{p(T,mu)}
    \\
    \hspace*{-2.5mm}
    \kappa(T)&=1+\Big(T\sfrac{\d}{\d T}\ln n(T,\mu(T))\Big)^{-1}
    \hspace*{-3mm}
    \label{kappa(T)}
  \end{align}
\end{subequations}
and  ~$\mu_{\rm NR}^{~}\!=\mu-m$.~  Here  $F^{\,\rm  eq}$ denotes  the
relativistic thermal  \mbox{single-particle} occupation  with pressure
$p(T,\mu)$ and relativistic adiabatic  index $\kappa(T/m)$, the latter
displayed  for various  masses in  Fig.\;\ref{fig-kappa-adiabat}.  The
under-braced   part   in   Eq.\;(\ref{N(tau)})  specifies   the   {\em
  non-relativistic}\,  (NR)  part  of the  densities  $n(T,\mu)$  with
dimensionless  {\em fugacity}  (or  chemical  activity) $f_{\rm  eq}$.
Thereby   $\lambda_{\rm  rel}(T/m)$   is  a   convenient  relativistic
correction factor caused by the non-Gaussian forms of the relativistic
\pbox{$F^{\,\rm eq}(\k)$} with \pbox{$\k\!\in\RR^3$} and approximation
(\ref{lambda}) valid within 0.5\% precision till $T\!<2m$.

In each  local rest-frame entropy conservation  together with particle
number conservation further require
\begin{subequations}
  \label{s-entropy}
  \begin{align}
    \label{s=(1+TdT)n}
    \hspace*{-10mm}
    \s(T,\mu)&V(T)
    =\Big(1+\msbx{\!T\sfrac{\partial}{\partial T}}\Big)\,n(T,\mu)V(T)
    \\
    &= \underbrace{n(T,\mu)V(T)}_{N=const}\Big\{\!
    \underbrace
    {\!\sfrac{\mu_{\rm NR}^{~}}{T}+\!\sfrac52
      +\phi(T/m)\!}_{\s(T,\,\mu)/n(T,\,\mu)=const}\Big\},
    \hspace*{-2mm}
    \label{s-entropyNR}
    \\
    \hspace*{-5mm}
    \sfrac{\mu_{\rm NR}^{~}(T)}{T}&=\ln f(T)=\ln f(T_0)
    -\phi\Big(\!\sfrac{T_0}{m}\Big)\!+\phi\Big(\!\sfrac{T}{m}\Big),
    \hspace*{-4mm}
    \label{adiabat-mu(T)}
    \\
    \hspace*{-5mm}
    \phi\Big(\!\sfrac{T}{m}\Big)\!:\!
    &=\msbx{T\sfrac{\partial}{\partial T}} \ln\lambda_{\rm rel}(T/m)
    =\frac{\frac{2T}{m} +\frac{T^2}{m^2}}
    {1\!+\!\frac{2T}{m}+\frac{T^2}{2m^2}}.
    \hspace*{-3mm}
    \label{phi-corrNR}
  \end{align}
\end{subequations}
Relation (\ref{adiabat-mu(T)})  defines the  adiabatic courses  of the
chemical potentials with  relativistic corrections $\phi(T/m)$ defined
in  (\ref{phi-corrNR}).   The  corresponding  leading  terms  are  the
standard    NR    forms,     like    the    Sackur--Tetrode    formula
\mbox{\cite{:O-Sackur1911,:H-Tetrode1912}}   for   the   \mbox{single}
particle entropy (\ref{s-entropyNR}).

The decoupling probability attains its maximum at
\begin{subequations}
  \label{dec-max}
  \begin{align}
    \label{dec-max-a}
    \Big[\sfrac{\d}{\d\tau}\Gamma(\k,\tau)
      +\Gamma^2(\k,\tau)\Big]_{\tau_{\rm max}^{~}}=&\;0.~~~~~
  \end{align}
  The  following    toy-model
  \begin{align}
    \label{t_max}
    \Gamma(\tau)=
        \Gamma_0\Big(\!\sfrac{\tau_0^{~}}{\tau}\Big)^{\!3}
    &\Rightarrow\;\Gamma^2({\tau_{\rm max}^{~}})\;
    ={\sfrac{3^{~}}{{\Gamma_0^{~}}\tau_0^3}}
    =\sfrac{1}{\tau_{\rm max}^2},
    \\
    \label{P_dec(p,tau)}
    \;\;\Delta\tau^{\rm dec}
    &\approx\sfrac{1}{{\cal P}^{\rm dec}(\tau_{\rm max}^{~})}
    \approx\sfrac{\rm e}{\Gamma(\tau_{\rm max}^{~})}.
    \hspace*{9.5mm}
  \end{align}
\end{subequations}
clarifies       the       dependencies      on       the       initial
\mbox{$\Gamma_0:=\Gamma(\tau_0)$}
\,\mbox{\cite{Knoll:2008sc,Mukhanov:2005sc}} with the {\em simple rule
  of  thumb}\,: The  stronger $\Gamma_0$,  the later  and broader  the
decoupling window.

On  the  {left  side}  of   each  curve  in  Fig.\,\ref{fig-Pdec}  the
interaction rates  along the  escape paths are  such high,  that those
particles cannot  undisturbed decouple:  \ie the  exponential survival
probabilities in (\ref{P^dec}) tend to zero:\\
\centerline{{\em  Those  parts  of  the  medium  are  opaque  for  the
    detectors}\,!}

The general  rule along  the escape  paths: $1/\Gamma$  determines the
mean free  lifetime to  remain in  the present  mode.  In  the Quantum
sense, \cf  \cite{Knoll:2008sc,Ivanov:1999tj}, the  spectral functions
are   correspondingly  broadened\footnotemark[5],   \mbox{unless}  the
particles  can freely  decouple.  With  decreasing $\Gamma(\tau)$  the
medium  becomes  successively  transparent   and  the  ${\cal  P}_{\rm
  dec}(\tau)$  \mbox{attain} forms  similarly  to  those displayed  in
Fig.\,\ref{fig-Pdec}.

For  multi-particle  systems  total  entropy along  with  the  overall
conservation laws and detailed balance have to be fulfilled, which may
lead to  complicated evolutions.  Still, the  above decoupling concept
does  allow for  {\em  entirely \mbox{analytical}  analyses} with  far
reaching consequences.\\[1mm]
\centerline{\bf ------------ $\ast$ ------------}

Alternatively, the continuous decoupling mechanism can be demonstrated
in transport simulations, which sample  the last interaction events of
each      particle      type,      \,\cf      \eg      Fig.\,6      in
Ref.\,\cite{Oliinychenko:2018ugs},    which    correspond    to    the
\mbox{$N(\tau)$}-distributions defined in Eq.\,(\ref{dec-rate-a}).
The  individuality of  decoupling processes  was already  demonstrated
about  two decades  ago  in various  transport  calculations, \cf  \eg
\mbox{\cite[Fig.\,4]{vanHecke:1998yu}},
\cite[Fig.\,23]{Forster:2007qk},      \cite[Fig.\,11]{Akkelin:2008eh},
\cite[Fig.\,2]{Pratt:2008qv}.

\section{Evolution steps of central nuclear collisions}
\label{Evolution_Steps}%
Here the essential  evolution steps from the hadronization  stage until the
late decoupling of the composite nuclei,  the main focus of this note,
are briefly summarized.

\subsection{The Hadronization Stage}
  
This  stage  concerns the  conversion  from  the  {QCD phase}  to  the
{hadronic phase}.   Also this  phase-transition is  continuous, during
which  {all hadrons}  are produced  and thereby  continuously decouple
from the  QCD phase in  the here  addressed sense.  In  ``our'' Flavor
Kinetic  model\,\cite{Barz:1988md,Barz:1990qy} the  decoupling process
lasted   about  5\,fm/c   or  even   shorter,  \cf   \eg  Fig.\,1   in
\cite{Barz:1988md}. The  created hadrons were produced  in approximate
chemical equilibrium\footnote{For  years our Flavor Kinetic  model was
the only  phase-transition model that {\em  preserved detailed balance
  and the Onsager relations}\,,  an essential requirement for reliable
predictions, thanks to  a recommendation by Gordon  Baym, to formulate
the rate equations  be driven in terms of  {\em chemical potentials}\,
rather than by the commonly used particle densities.}.
 
{\bf  Mesons}, that  couple weakly  to the  hadronic medium,  have the
chance  to be  {\em early  messengers of  the hadronization  stage}\,.
Thereby   heavy  mesons,   like  charm   mesons,  are   burdened  with
mass-thresholds, \eg \mbox{$\propto\exp(-m_{c\Bar  c}/T)$}, which drop
fast \mbox{with $T$}.

{\bf  Baryon--anti-baryon  annihilation:}  At collider  energies  this
process   is   shown    by   data   to   quickly    cease,   \cf   the
in-medium\!\footnote{The term ``in-medium''  is used for calculations,
which determine  the in-medium  properties \eg till  some temperature,
without explicitly caring  about the here discussed  decoupling of the
observed particles.}  results\,\cite{Pan:2014caa}, such that from then
on  baryons and  anti-baryons  evolve essentially  decoupled from  one
another     with      individual     adiabatic     \mbox{$\{T,\mu_{\rm
    B}(T)\}$}-courses.   For  reactions  at lower  energies,  see  for
instance         the         in-medium         calculations         of
Refs.\,\cite{Rapp:2000gy,Greiner:2000tu}.

\subsection{Collective Flow}
\label{Sec-Coll-Flow}
Collective flow  is generated from  the moment of  highest compression
on.   Depending on  bombarding energy  this may  include its  creation
during the  QCD-phase and the entire  hadronization process presumably
then generating  the major amount of  flow\,\cite{Barz:1988md}.  Since
pressure gradients  cease fast  with the dilution  of the  system, the
collective  flow may  quickly  {\em freeze in},  in accordance  with
introductory Note\;\ref{Freeze-in}.

A confirming sign  of collective flow is that  the mean-square momenta
$\big<k^2\big>$ of the observed particles' momentum spectra scale with
the square of their masses, \ie $\big<k^2\big>\propto m^2$ such that
{$\big<v^2\big>\approx\big<k^2/(m^2+k^2)\big>\approx
  const.$}

The mean transverse  flow velocity can be deduced \eg  from the maxima
of the transverse momentum spectra, which scale linear in $p_\perp$ at
low     momenta.      For     all     NR-particles     measured     in
Ref.\,\cite{ALICE:2023qyl}   these   maxima   are   occurring   around
\pbox{$\big<p_{\perp}\big>\approx m$} determining  the mean transverse
collective   flow    velocity   to   \pbox{$\big<v_{\perp}\big>\approx
  0.7\,{\rm c}\,$}.

Assuming  \eg  a  flow saturation  around  \mbox{$T\approx  150\,$MeV}
allows  to estimate  the maxima  of the  decoupling windows.   For the
model  calculations of  Ref.\,\cite[Fig.\,6]{Oliinychenko:2018ugs} the
maximum of  Deuterons can be  determined to occur at  around 1/8$^{\rm
  th}$   of    the   initial    density   within   a    $T$-range   of
\mbox{$[70\rightarrow\!   40]$\,MeV},   here  roughly   analyzed  with
adiabatic     indices     within    the     range     \mbox{$\kappa\in
  [1.45\rightarrow\!5/3 ]$}, \cf Fig.\,\ref{fig-kappa-adiabat}.
\subsection{The Pion-Nucleon-Delta Entanglement}
\label{Sec-pi-N-Delta}

\begin{figure}[t]
  \begin{center}
    \caption{}
    \label{fig-pi-N-Delta}
  \begin{picture}(70,54)
    \put(-1,-4.5){\includegraphics[width=0.415\textwidth]{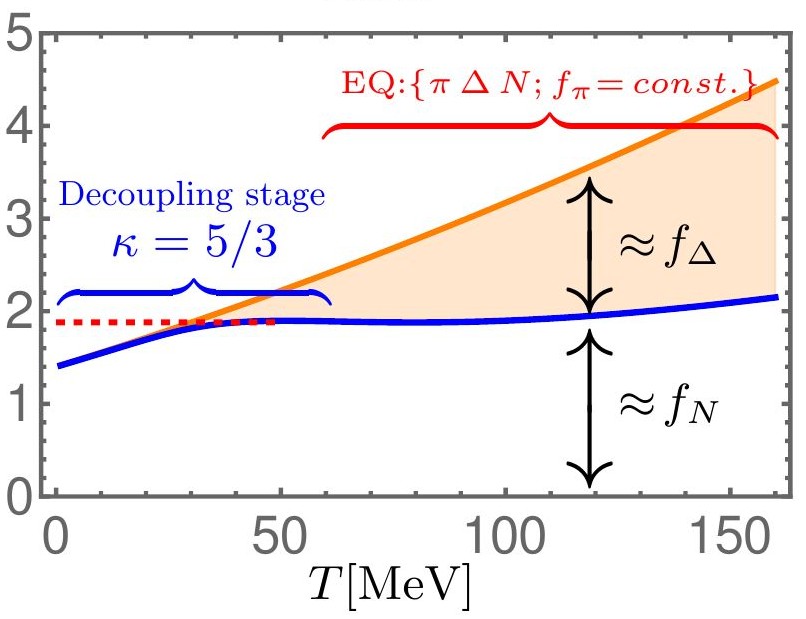}}
  \end{picture}
  \end{center}
  \begin{trivlist}
  \item {\bf Pi-{\em N}-Delta Entanglement} here estimated in a simple
    equilibrium model  with constant pion fugacity  $f_\pi$.  It shows
    the two  stages of the dynamics.   Above \mbox{$T\approx 60$\,MeV}
    the $\pi N\Delta$-cycles hinder the formation of composite nuclei;
    \mbox{below} this temperature pions  decouple from the baryons and
    the  baryon   evolution  reaches  the  NR-decoupling   stage  with
    $\kappa\approx5/3$.
  \end{trivlist}%
\end{figure}

\indent   A  special   role  plays   the  overwhelming   entourage  of
\mbox{pions}.   With \mbox{$p$-wave  } cross-sections\,  of 16  fm$^2$
\cite{:DeltaResonance1952}    each    pion    is   eager    to    form
\mbox{$\Delta$-resonances}  with the  nucleons in  continuous creation
and  decay cycles  till approximately  \mbox{$T\approx 60\,$MeV},  \cf
Fig.\,\ref{fig-pi-N-Delta}                   and                   \eg
\mbox{Ref.\,\cite[Figs.\;6$\;\&\;$8]{Oliinychenko:2018ugs}}.     These
short-term  cycles will  seriously  \mbox{obstruct}  the formation  of
bound nuclei.
\unitlength1.mm

A particular influence of pions  on the nuclear bound state abundances
is    shown    \eg   by    the    recent    transport   approach    of
Ref.\;\,\mbox{\cite[Fig.\,3c]{Sun:2022xjr}}  for  the  Triton  yields.
The relatively early stabilization of  the baryon abundances and their
ratios were shown in transport models with point-like particles \eg in
Refs.\,\cite{Staudenmaier:2021lrg,   Vovchenko:2019aoz,  Neidig_2022},
indicating   their  early   freezing-in  already   during  this   high
temperature phase in the sense of introductory Note\;\ref{Freeze-in}.

Still,   sooner   or    later   along   with   the    cease   of   the
\mbox{$\Delta$-resonance cycles} (the precise occurrence is relatively
unimportant) the pions will no  longer significantly interact with the
baryons.   This separates  the pion  dominated high  temperature phase
with its own  adiabatic features from the  subsequent decoupling stage
of  the  baryons, which  from  then  on  will  evolve with  their  own
adiabatic  courses, \cf  Fig.\,\ref{fig-pi-N-Delta}, and  as discussed
below in Sec.\,\ref{Sec-L-Nuc}.
\subsection{Bound States in Matter}
\label{Sec-Dissolution}
Bound   states   are   the   most   vulnerable   objects   in   matter
environments. First, their  sizes cause large cross  sections with the
surrounding  particles.   The   correspondingly  large  damping  rates
$\Gamma$ generally push their  decoupling windows towards lower matter
densities than those of their respective constituents.

Secondly, the surrounding matter  influences the binding properties of
bound  states.   In the  here  discussed  low energy  nuclear  physics
context    the    problem    was   investigated    by    our    German
colleagues\,\mbox{\cite{ROPKE1982536,  ROPKE1983587}}  accounting  for
Pauli-blocking     effects.     The     relativistic    mean     field
considerations\,\cite[Fig.\;1]{Typel:2009sy},   however,   showed   no
significant effect along the here relevant adiabatic NR-parkour.

Still, the  phenomenon is  by far  more general:  in dense  matter the
bound-state wave-functions can no  longer spatially extend to infinity
but are rather spatially  restricted.  Such spatial restrictions cause
a  spatial squeezing  of  the bound-state  wave-functions inducing  an
increase of the kinetic energy part of their total bound-state energy.
The latter has the effect, that above a certain density the states are
no  longer  bound.  Together  with  the  collisional broadening  these
states then form resonances,  whose continuous spectral functions then
have  pole  positions  above  the bound-states'  nominal  masses.   In
particular  such geometric  effects can  \eg significantly  affect the
Hypertriton's           halo-tail           for           temperatures
$T\ge5$\,MeV\,\footnote{Thanks  to  both  a geometrical  and  a  rough
Beth-Uhlenbeck     estimate\,\cite{Beth:1937zz}     exploiting     the
entanglement  between bound  state properties  and phase-shifts  by my
local colleague B.  Friman.}.  Again the correct individual values are
less important.

The on-shell formation of  bound states requires a \mbox{simultaneous}
interaction with a third party,  here properly provided by the BE-rate
$\Gamma(\tau)F(\x,\k,\tau)$ in Eqs.\,(\ref{dec-rate}).

Classical transport calculations can still approximately determine the
integrated  spectral strength  of  bound states.   Thereby their  real
bound-state  nature will  be  restored  at the  moment  of their  last
interaction   before   reaching   infinity,   as   \eg   recorded   in
Ref.\,\cite[Fig.\,6]{Oliinychenko:2018ugs}.

\subsection{Decoupling of light Nuclei}
\label{Sec-L-Nuc}

The  above conceptual  considerations in  Sects.\,\ref{Sec-pi-N-Delta}
and \ref{Sec-Dissolution} suggest that  the decoupling of light nuclei
has    to   occur    by   far    later   than    so   far    generally
anticipated\,\cite{arXiv:1710.09425},   namely   along  an   adiabatic
course, where  the pions  essentially do no  longer interact  with the
(anti-)baryons.   Rather   than  using  the  thermal   fit  parameters
$\big(T_0,\mu_{\rm B}(T_0)\big)$  let us  focus on  the fact  that the
fitted {\em fugacities  of all particles} can be interpreted  as to be
\mbox{\em dynamically constant}\,!

Upon  looking at  the  comparison of  the  experimentally observed  so
called  \mbox{primordial  abundances} of  the  light  nuclei, \cf  the
dashed   line   in~Ref.\,\mbox{\cite[Fig.\,2]{arXiv:1710.09425}}   one
states an  approximate linear  behavior between  the logarithm  of the
individual fugacities $f(m)$  versus mass $m$.  With  reference to the
nucleon with mass $m_{\rm N}$ the observed data show
\begin{subequations}
  \label{dec-NR}
  \begin{align}
    f(m)\approx f(m_{\rm N})^{m/m_{\rm N}} .
    \label{ln-f(T,mu;m)}
  \end{align}
  How can the above fugacity-systematic ever comply with the \mbox{\em
    individuality}  of a  continuous  decoupling  process along  which
  temperatures continuously drop?

  Well,  this  Gordian  knot   was  cut  200  years  ago\footnote{With
  discoveries of  those thermodynamic  processes, investigated  \eg by
  S.\,D.\;Poisson    (1823),    N.\,L.\,S.\;Carnot\,\cite{:Carnot1824}
  (1824)  and  others,  whose   properties  were  classified  as  {\em
    adiabatic}\,  a  few decades  later.}.   Therefore  my {\em  sole}
  interpretation and  conclusion is, that these  light nuclei decouple
  by  {far   later},  \cf  Fig.\,\ref{fig-pi-N-Delta},   along  common
  NR-adiabatic courses.   As a~key feature  in this context,  not only
  \mbox{\em  entropy}\, and  {\em  abundances}\,  of NR-particles  are
  preserved along  NR-adiabates but  also their {\em  fugacities}, \cf
  Eqs.\,(\ref{s-entropy})    with    NR-property
  \mbox{$\phi(T/m)\equiv 0$} in (\ref{adiabat-mu(T)}).

  At the here discussed decoupling stages the inter-particle distances
  between the  baryons are such  large that  an ideal gas  equation of
  state, dominated by the nucleons,  is well justified for all baryons
  with \mbox{$\kappa=5/3$} valid for 3 translation degrees of freedom.
  Their  NR-evolutions  then  proceed  in  accordance  with  Poisson's
  adiabatic laws  (1823). The  corresponding decoupling yields  of the
  (composite) baryons (\ref{dec-rate-a}) then simplify to
  \begin{align}
    \hspace*{-3mm}
    N^{\rm
      dec}(m)&=\!\int\!\!\sfrac{\d^3x\d^3k}{(2\pi\hbar)^3}\!
    \;\e^{-\varepsilon_{\rm NR}(k)/T}\!
    \underbrace
        {\int_{0}^\infty\!\!\!\d\tau\,{\cal P}^{\rm dec}(\tau) f(m)}
        _{\equiv\; f(m)}\!
        \label{f(m)P^dec}
        \\[-2mm]
        \label{m^3/2f(m)}
    &=\!\underbrace{\Big(\!\sfrac{mT}{2\pi\hbar^2}\Big)^{\!3/2} V(T)}
        _{\rm adiabaticly\; constant}\! f(m)\propto m^{3/2} f(m),
    \\
    \label{mu(T,m)}\hspace*{-2.5mm}
    \ln f(m)&=\mu_{\rm NR}^{~}(T,m)/T,
    ~~\varepsilon_{\rm NR}^{~}(k)=k^2/(2m)
\end{align}
\end{subequations}
 with    Poisson's   \mbox{adiabatic}    relation   under-braced    in
 line~(\ref{m^3/2f(m)}).  The latter further determines the correct NR
 mass-dependence  as \mbox{$N_m^{\rm  dec}\propto m^{3/2}  f(m)$}, \cf
 Eq.\,(\ref{dec-rate-a}). For  the above steps both,  the constancy of
 the       $f(m)$,      \cf       Eq.\;(\ref{adiabat-mu(T)})      with
 \mbox{$\phi(T/m)\equiv    0$},   together    with   unity    integral
 (\ref{P^dec-unity}) over ${\cal P}^{\rm dec}(\tau)$ were used.

The fact  that the mass  systematic of the  measured \mbox{fugacities}
(\ref{ln-f(T,mu;m)}) could  be fitted by  one common set  of $(T,\mu)$
further  implies  that  nucleons  and composite  nuclei  are  in  {\em
  chemical  equilibrium}  to one  another.   The  NR-constancy of  all
$f(m)$  further  confirms this  feature  maintained  along the  entire
decoupling courses.

Thus, it is {\em neither} important,  when and during which later time
span the measured composite baryons  decouple!  {\em Nor} is there any
need to  determine their  individual decoupling  windows: The  Laws of
Nature    will    just    properly    care    in    accordance    with
Eqs.\;(\ref{f(m)P^dec})\,\ff!

The  presented  analysis  uses  independent  particle  concepts  under
circumstances, where they are well  at place.

Notably,  neither  the  individual  decoupling windows  nor  thus  the
decoupling  values of  $T$ or  adiabatic index  $\kappa$ are  directly
experimentally measurable.

{\em A  final clarification:}\, All  in-medium properties such  as the
above  discussed   thermo-chemical  equilibration  are  by   no  means
conceptually affected by the  decoupling features respectively used on
the    \mbox{\em     r.h.s.}     of     Eqs.\;(\ref{dec-rate-a})    or
(\ref{f(m)P^dec}).  The  latter are encoded via  ${\cal P}^{\rm dec}$,
whose  individual   decoupling  windows  are   solely  retrospectively
clarified  at ``\mbox{\em  post-completion}'' of  the system's  entire
global  evolution, \ie  at $\tau\rightarrow\infty$  via the  relations
discussed in Sect.\;\ref{DecPicture}.

\subsection{Entropy}
\label{Sec-Entropy}
In my view the here discussed experiments preferentially determine the
{entropy per particle} rather than the hadronization temperature.

Since for NR-particles relativistic  correction $\phi$ vanishes, their
entropy   per  particle   can  be   obtained  from   the  under-braced
Sackur-Tetrode      factor\,\cite{:O-Sackur1911,:H-Tetrode1912}     in
Eq.\,(\ref{s-entropyNR})  defining  the   running  chemical  potential
(\ref{adiabat-mu(T)}) for non-relativistic baryons.  This then implies
the  constancy of  their $f(m)$,  \cf Eq.\,(\ref{adiabat-mu(T)}),  and
thus   allows    to   determine    the   entropy   of    any   created
\mbox{(anti-)}baryon.  This strategy agrees  with the one suggested by
Siemens and Kapusta\,\cite{Siemens:1979dz}  concerning the $d/p$-ratio
some  40  years  ago,  here presented  in  the  continuous  decoupling
picture.

The  determination of  the entropy  of  the dominating  pions is  more
subtle   and   definitely   model  dependent.    As   {Nambu-Goldstone
  bosons}\,\mbox{\cite{Nambu:1960,Goldstone:1961}}  their   number  is
approximately  conserved, \cf  \eg \cite{Pan:2014caa}.   Assuming that
only  a minority  of  pions is  involved in  \mbox{$\Delta$-formation}
cycles   allows   to   obtain   their   entropy   per   particle   via
Eq.\,(\ref{s-entropyNR})  through the  experimentally determined  pion
fugacity  $f(m_\pi,T_\pi)$ still  for  an appropriately  to be  chosen
temperature $T_\pi$.

The  total  entropy,  which  discloses  information  about  the  early
situation  of   entropy  saturation  around  the   moment  of  highest
compression, is  then determined through  the sum of the  baryon parts
(dominated by nucleons) plus the meson part (dominated by pions).

\section{Summary}
\label{Summary}%
The  here   presented  analytical  considerations   resolve  dynamical
inter-plays  that  can  barely   be  disentangled  by  mere  numerical
simulations. They  explain the  for a long  time puzzling  features of
freeze-out and  decoupling mechanisms  of central high  energy nuclear
collisions.   The  success  rests  on the  general  concept  that  any
transition process is {\em individual and has its own time frame}.

None the  less certain observables,  like the abundances  of particles
and  their  ratios  \cf  \eg  \mbox{Refs.\,\cite{Staudenmaier:2021lrg,
    Vovchenko:2019aoz,  Neidig_2022}},  can already  approximately  be
stabilized   relatively  early,   here  referred   to  as   \mbox{{\em
    freezing-in}}.  Since  pressure gradients quickly cease,  also the
collective  flow field  may stabilize  correspondingly, this  way also
predetermining  the  momentum spectra  in  the  sense of  introductory
Note\;\ref{Freeze-in}.

Still,  all  experimentally  observed   reaction  products  will  {\em
  \mbox{definitely} decouple later}.   Thereby these processes resolve
the  latest in-medium  situations, which  the observed  particles have
encountered.   They determine  their  abundances  and momenta  notably
during their respective decoupling  stages in a model-independent way.
Any  extrapolation  to  earlier  evolution  situations  would  require
appropriate model estimates.

How is it  then possible, that all abundances can  simply be described
by solely  two parameters,  although the decoupling  processes proceed
individually?

This rests on a well hidden intricate interplay between the early high
temperature phase  dominated by  pions and  the later  low temperature
evolution. During the latter the  here addressed baryon sector evolves
essentially  decoupled from  the  pions with  its  special well  known
adiabatic properties of  non-relativistic particles, namely preserving
their  fugacities along  the  entire  decoupling \mbox{parkour}.   The
latter assures  that the particular decoupling  yields are independent
of the temporal appearances and widths of the corresponding decoupling
windows.   Therefore  the  formalism  gives  a  physically  sound  and
rigorous  explanation of  the observed  wealth of  data. It  naturally
provides  the chance  for the  undisturbed production  of the  faintly
bound \mbox{(Anti-)}Hypertritons.  The latter rests on a very late and
sufficiently dilute  local decoupling scenario with  local bound-state
formation then within the realm of standard low energy nuclear physics
even at sub-picometer scales.

As  explained in  Sect.\,\ref{Sec-Entropy},  the method  allows for  a
robust determination of the entropy content carried by the baryons and
anti-baryons.  Concerning the high temperature  sector there is a need
to clarify the  non-equilibrium aspects of pions and  mesons along the
entanglement  process   sketched  in   Sect.\,\ref{Sec-pi-N-Delta}  by
appropriate  model  studies,  \eg   \mbox{using}  the  techniques  and
decoupling tools of Sect.\;\ref{DecPicture}.

In  1907 G.\;W.\;Lewis\,\cite{Lewis:1907}  has  shown that  fugacities
({\em activities} in~chemistry) are measurable obsevables, \eg through
a thermo-chemical  contact with a calibrated  \mbox{reference} system.
In this context, the fugacities  determined in these central collision
experiments   are  then   straight   messengers   of  the   particles'
\mbox{physical  in-medium} properties  during their  proper individual
decoupling stages.\\[-8mm]
\section*{Declaration of competing interests}
As senior  scientist I  declare to have  no known  competing financial
interests  or personal  relationships that  could have  influenced the
work reported in this paper.\\[-8mm]
\section*{Acknowledgment}
I  thank  my  (local) colleagues,  A.\;Andronic,  P.\;Braun-Munzinger,
B.\;Friman, C.\;Greiner, H.\;van\;Hees  and J.\;Wambach for clarifying
\mbox{discussions},   suggestions   and  encouragements   on   various
occasions.
Further thanks  address S.\,Pratt  and Y.\,Sinyukov,  who
emphasized the importance of adiabatic behaviors in the context of HBT
and  the  here  addressed   decoupling  \mbox{phenomena}  during  some
discussions with me already more than a dozen years ago.

\bibliographystyle{apsrev4-1}
\bibliography{References}
\end{document}